\documentclass[useAMS,usenatbib,usegraphicx]{mn2e}
\usepackage{color}

\title[New thresholds for PBH formation]{New thresholds for Primordial Black Hole formation during the QCD phase transition}
\author[J. L. G. Sobrinho, P. Augusto and A. L. Gon\c{c}alves]{J. L. G. Sobrinho$^{1,2}$\thanks{E-mail: sobrinho@uma.pt (JLGS)}, P. Augusto$^{1,2}$\thanks{E-mail: sciman@med.up.pt (PA)}, and A. L. Gon\c{c}alves$^{1}$\thanks{E-mail: angel@uma.pt (ALG)}\\
$^{1}$Faculdade de Ci\^{e}ncias Exatas e da Engenharia, Universidade da Madeira, Caminho da Penteada, 9000-390 Funchal, Portugal\\
$^{2}$Centro de Astronomia e Astrof\'{i}sica da Universidade de Lisboa, Tapada da Ajuda, Edif\'{i}cio Leste, 2$^{\circ}$ piso, 1349-018 Lisboa, Portugal}

\begin{document}

\date{Accepted xxxx. Received xxxx; in original form xxxx}

\pagerange{\pageref{firstpage}--\pageref{lastpage}} \pubyear{2013}

\maketitle

\label{firstpage}

\begin{abstract}
Primordial Black Holes (PBHs) might have formed in the early Universe as a consequence of the collapse of density fluctuations with an amplitude above a critical value $\delta_{c}$: the formation threshold.
Although for a radiation-dominated Universe $\delta_{c}$ remains constant, if the Universe experiences some dust-like phases (e.g. phase transitions) $\delta_{c}$ might decrease, improving the chances of PBH formation. We studied the evolution of  $\delta_{c}$ during the QCD phase transition epoch within three different models: Bag Model (BM), Lattice Fit Model (LFM), and Crossover Model (CM). We found that the reduction on the background value of $\delta_{c}$ can be as high as $77\%$ (BM), which might imply a $\sim10^{-10}$ probability of PBHs forming at the QCD epoch.
\end{abstract}

\begin{keywords}black hole physics -- cosmology: early Universe -- cosmology: inflation\end{keywords}

\section{Introduction}
\label{sec:Introduction}

Primordial Black Holes (PBHs) may have formed in the early Universe as a consequence of the collapse of density fluctuations \citep[e.g.][]{1971MNRAS.152...75H, 1974MNRAS.168..399C, 1975ApJ...201....1C, 1979A&A....80..104N, 2001IJMPD..10..927P, 2010RAA....10..495K}. They might even be directly detectable within our neighboorhood \citep[][]{2014MNRAS.441.2878S}. During inflation, fluctuations of quantum origin are  stretched to scales much larger than the cosmological horizon~$R_H$ at the time $t$ when they were produced ($R_{H}(t)=c/H(t)$, with $H(t)$ the Hubble parameter). Once a physical wavelength becomes larger than $R_{H}$, it is causally disconnected from physical processes. The inflationary era is followed, respectively, by radiation-dominated  and matter-dominated epochs during which these fluctuations can re-enter the cosmological horizon~\citep[e.g.][]{2006ARNPS..56..441B}. For a given physical scale $k$, the horizon crossing time $t_{k}$ (i.e. the instant when that scale re-enters $R_{H}$) is conventionally defined by \citep[e.g.][]{2003PhRvD..67b4024B,2002PhRvD..65b4008B}
\begin{equation}
\label{horizon crossing}
ck=a(t_{k})H(t_{k})
\end{equation}
where $a(t)$ represents the scale factor. The collapse that gives rise to the formation of a PBH is now possible but only if the amplitude of the density fluctuation ($\delta$) is larger than a specific threshold value $\delta_{c}$. In this case the expansion of the overdense region will, eventually, come to a halt, followed by its collapse. 
The majority of the PBHs formed at a particular epoch have masses within the order of the horizon mass, $M_{H}$, at that epoch \citep[e.g.][]{2003LNP...631..301C} given by
\begin{equation}
\label{horizon-mass}
M_{H}(t)\sim 10^{15}\left(\frac{t}{10^{-23}\mathrm{~s}}\right)\mathrm{~g}.
\end{equation}
However, in the case of perturbations with $\delta$ only slightly larger than the critical value, $\delta_c$, the PBH masses rather obey the scaling law 
\citep[][]{1999PhRvD..59l4013N}
\begin{equation}
\label{scaling-law}
M_{PBH}\propto M_{H}\left(\delta-\delta_{c}\right)^{\gamma}
\end{equation}
where $\gamma\approx 0.36$ in the case of a radiation-dominated Universe. This power law scaling has been found to hold down to $(\delta-\delta_{c})\sim 10^{-10}$ \citep[][]{2009CQGra..26w5001M,2013CQGra..30n5009M}.

The probability that a fluctuation crossing the horizon at some instant $t_{k}$ has of collapsing and forming a PBH can be written as \citep[e.g.][]{green2015}
\begin{equation}
\label{Bringmann_et_al_2001_eq5}
\beta(t_{k})=\frac{1}{\sqrt{2\pi}\sigma(t_{k})}\int_{\delta_{c}}^{\infty}
\exp\left(-\frac{\delta^{2}}{2\sigma^{2}(t_{k})}\right)d\delta
\end{equation}
where $\sigma^{2}(t_{k})$ is the mass variance at horizon crossing.

\citet[][]{1975ApJ...201....1C}, based on a simplified model of an overdense collapsing region, found $\delta_{c}=1/3$ for a radiation-dominated Universe. In more recent years the threshold $\delta_{c}$ has been extensively investigated by numerical simulations in PBH formation, in particular by numerically solving relativistic hydrodynamical equations \citep[see e.g.][]{2005CQGra..22.1405M}. Although \citet[][]{1999PhRvD..59l4013N} reported the value $\delta_{c}\simeq 0.7$ in the case of a radiation-dominated expanding Universe, this was later revised to $\delta_{c}\simeq 0.43 - 0.47$ \cite[][]{2005CQGra..22.1405M, 2007CQGra..24.1405P, 2009CQGra..26w5001M, 2013PhRvD..88h4051H, 2013CQGra..30n5009M}. 
Usually $\delta_{c}$ is constant throught the radiation-dominated epoch, the exception occurring during cosmological phase transitions, when the value of  $\delta_{c}$ decreases (as a consequence of the decrease of the sound speed). This is relevant, since a lower value of  $\delta_{c}$ favours PBH production \citep[e.g.][]{2003LNP...631..301C}.

The Standard Model of Particle Physics (SMPP) predicts: (i) the Electroweak (EW) phase transition at temperatures of $\sim 100\mathrm{~GeV}$, responsible for the spontaneous breaking of the EW symmetry \citep[e.g.][]{2006Natur.443..675A} and (ii) the Quantum Chromodynamics (QCD) phase transition at $170\mathrm{~MeV}$ \citep[e.g.][]{2002PhRvD..66g4507A, 2005PhRvD..71c4504B, 2006PhLB..643...46A} related to the spontaneous breaking of the chiral symmetry of the QCD when quarks and gluons become confined in hadrons. At very high temperatures ($T> 172.5\mathrm{~GeV}$) all the particles of the SMPP contribute to the effective number of degrees of freedom giving g(T)=106.75 \citep[e.g.][]{2006JPhG...33....1Y, sobrinho2011}. As the expansion of the Universe goes on, the temperature decreases and, by the temperature of the QCD transition, with the Universe consisting on a Quark-Gluon Plasma (QGP), we have $g_{QGP}=61.75$. At the end of the QCD transition, when the Universe becomes an Hadronic Gas (HG), we reach $g_{HG}=17.25$ \citep[e.g.][]{2012PhRvD..86a0001B}.

The aim of this paper is to study the behaviour of $\delta_{c}$ during the QCD phase transition. The paper is organized as follows: after reviewing, in Section \ref{sec:The early Universe}, some key aspects concerning the early Universe and the QCD phase transition we derive, in Section \ref{sec:delta_c QCD}, new results for $\delta_{c}$ considering three different models: the Bag Model (BM), the Lattice Fit Model (LFM), and the Crossover Model (CM). In Section \ref{sec:discussion}  we conclude with a general discussion.

We have considered, for the radiation-dominated phase of the Universe, the lowest accepted value of $\delta_{c}=0.43$, which corresponds to the Mexican-Hat perturbation profile, a very representative one.
Table \ref{tabela-inicial} (Section~\ref{sec:The early Universe}) sums up key observational and derived parameters that we used throughout this paper.

\begin{table*}
\caption[]{Observational and derived (the last three) parameters used in our calculations: {\bf (1)} nomenclature; {\bf (2)} description (HG --- Hadronic Gas; QGP --- Quark-Gluon Plasma); {\bf (3)} numerical value (BM --- Bag Model; LFM --- Lattice Fit Model); {\bf (4)} reference, if any: [1] \citet[][]{2014A&A...571A..16P}; [2] \citet[][]{2013ApJS..208...20B}; [3]~\citet[][]{2002PhRvD..66g4507A, 2005PhRvD..71c4504B, 2006PhLB..643...46A} [4] \citet[][]{2012PhRvD..86a0001B}.
\label{tabela-inicial}
}
\center
\begin{tabular}{lp{10cm}lc}
\hline
{\bf (1)} & {\bf (2)}  & {\bf (3)}  & {\bf (4)} \\
\hline
$t_{0}$ & Age of the Universe & $4.36\times10^{17}$~s & [1]\\
$T_{0}$ & Cosmic Microwave Background Temperature & 2.72548~K & [2]\\
$H_{0}$ & Present day value of the Hubble Parameter $H(t)$& $67.3\mathrm{~kms}^{-1}\mathrm{Mpc}^{-1}$ & [1]\\
$\Omega_{\Lambda}$ & Dark energy density parameter & 0.685 & [1] \\
$t_{\Lambda}$ & Age of the Universe at matter-$\Lambda$ equality (when the expansion starts to accelerate); equals $2/(3H_0)$ & $3.06\times10^{17}\mathrm{~s}$ & ---\\
$z_{eq}$ & Redshift at matter-radiation equality & 3391 & [1]\\
$t_{eq}$ & Age of the Universe at matter-radiation equality (from equation (\ref{scale factor matter}), considering that $a_{m}(t_{eq})=(1+z_{eq})^{-1}$) & $2.37\times10^{12}\mathrm{~s}$ & --- \\
$T_{c}$ & The temperature of the Universe at the QCD phase transition & 170~MeV & [3]\\
$g_{QGP}$ & Degrees of freedom for the QGP Universe & 61.75 & [4]\\
$g_{HG}$ & Degrees of freedom for the HG Universe & 17.25 & [4]\\
$t_{+}$ & Age of the Universe at the end of the QCD phase transition & $1.2\times10^{-4}\mathrm{~s}$~(BM, LFM)  & Section \ref{sec:delta_c QCD}\\
$t_{-}$ & Age of the Universe at the begining of the QCD phase transition  & $6.2\times10^{-5}\mathrm{~s}$~(BM)  & Section \ref{sec:delta_c QCD}\\
 & & $9.5\times10^{-5}\mathrm{~s}$~(LFM) & Section \ref{sec:delta_c QCD}\\
\hline
\hline
\end{tabular}
\end{table*}

\section{The early Universe}
\label{sec:The early Universe}

\subsection{The scale factor}

We appear to live in a flat, homogeneous and isotropic expanding Universe (at scales $>$~100~Mpc) well described by the Friedmann-Lema\^{i}tre-Robertson-Walker (FLRW) metric where the \emph{scale factor}, $a(t)$, describes the time dependence of the Universe geometry: $a(t)\propto t^{1/2}$ for radiation domination, $a(t)\propto t^{2/3}$ for matter domination, and $a(t)\propto \exp(\sqrt{\frac{\Lambda}{3}}ct)$ for dark energy domination, where $\Lambda=3H_{0}\Omega_{\Lambda}c^{-2}$ is the cosmological constant \citep[e.g.][]{2012PhRvD..86a0001B} --- see Table~\ref{tabela-inicial}. A common normalization of the FLRW metric  defines the scale factor equal to unity at the present time \citep[$a(t_{0})=1$; e.g.][]{1993PhR...231....1L}, yielding, for a Universe dominated by a positive cosmological constant (cf. Table~\ref{tabela-inicial}) 
\begin{eqnarray}
\label{scale factor dark energy}
a_{\Lambda}(t)=\exp\left(c\sqrt{\frac{\Lambda}{3}}(t-t_{0})\right),&  t_{\Lambda}\leq t\leq t_{0} \: .
\end{eqnarray}
For the matter-dominated Universe we have
\begin{eqnarray}
\label{scale factor matter}
a_{m}(t)=a_{\Lambda}(t_{\Lambda})\left(\frac{t}{t_{\Lambda}}\right)^{2/3}, \:\:  t_{eq}\leq t\leq t_{\Lambda} \: .
\end{eqnarray}
Between the end of inflation at $t=t_{e}\sim 10^{-33}\mathrm{~s}$; \citep[e.g.][]{1991ARAA..29..325N} and $t=t_{eq}=2.37\times10^{12}$~s the Universe was radiation-dominated since at the latter instant a radiation-matter equality is reached (Table \ref{tabela-inicial}). Exceptions to radiation-domination in the $t_{e}\leq t\leq t_{eq}$ time interval  might have taken place during cosmological phase transitions such as the QCD (when the Universe might have been dust-like). We, then, split the radiation-domination epoch into three parts. For the interval between the end of the QCD phase transition ($t_{+}$) and $t_{eq}$ we define 
\begin{equation}
\label{scale factor radiation late}
a_{rl}(t)=a_{m}(t_{eq})\left(\frac{t}{t_{eq}}\right)^{1/2}, \:\: t_{+}\leq t\leq t_{eq} \: .
\end{equation}
During the QCD phase transition we define
\begin{eqnarray}
\label{scale factor dust QCD}
a_{QCD}(t)=a_{rl}(t_{+})\left(\frac{t}{t_{+}}\right)^{n_{qcd}}, \:\: t_{-}\leq t\leq t_{+}
\end{eqnarray}
where $n_{qcd}=2/3$ if the Universe experiences a QCD dust-like phase or $n_{qcd}=1/2$ if the Universe continues to be radiation-dominated during that epoch. Finally, between the end of inflation ($t_{e}$) and the beginning of the QCD phase transition ($t_{-}$) we define
\begin{equation}
\label{scale factor radiation medium}
a_{rm}(t)=a_{QCD}(t_{-})\left(\frac{t}{t_{-}}\right)^{1/2}, \:\: t_{e}\leq t\leq t_{-} \: .
\end{equation}

\subsection{Primordial density fluctuations}

The simplest way to describe a classical fluctuation with amplitude $\delta$ is in terms of an overdensity or density contrast \citep[e.g.][]{2005CQGra..22.1405M}
\begin{equation}
\label{overdensity}
\rho=\overline{\rho}(1+\delta)
\end{equation}
where $\overline{\rho}$ represents the average cosmological density. In the unperturbed region we assume the FLRW metric and obtain \citep[][]{1975ApJ...201....1C}
\begin{equation}
\label{CardallFuller_eq4}
\left(\frac{da}{dt}\right)^{2}=\frac{8\pi G}{3}\overline{\rho}(t)a(t)^{2}
\end{equation}
while for the perturbed one we get \citep[][]{1975ApJ...201....1C}
\begin{equation}
\label{CardallFuller_eq6}
\left(\frac{ds}{d\tau}\right)^{2}=\frac{8\pi G}{3}\rho(\tau) s(\tau)^{2}-\Delta\epsilon
\end{equation}
where $\Delta\epsilon $ represents the corresponding curvature constant, $\tau$ is the proper time as measured by comoving observers and $s(\tau)$ plays the role of a scale factor for the perturbed region. Considering that, initially, the overdense region is comoving with the unperturbed background, we consider $\tau_{k}=t_{k}$, $s_{k}=a_{k}$ and $(ds/d\tau)_{k}=(da/dt)_{k}$ yielding, for $\Delta\epsilon $, the expression \citep[][]{1975ApJ...201....1C}  
\begin{equation}
\label{Delta_epsilon}
\Delta\epsilon =\frac{8\pi G}{3}a_{k}^{2}\left(\rho_{k}-\overline{\rho}_{k}\right).
\end{equation}
Inserting this into equation (\ref{CardallFuller_eq6}) and using  equation (\ref{overdensity}) we obtain
\begin{equation}
\label{CardallFuller_eq6a}
\left(\frac{ds}{d\tau}\right)^{2}=\frac{8\pi G}{3} \left(\rho(\tau) s(\tau)^{2}-\rho_{k}a_{k}^{2}\frac{\delta_{k}}{1+\delta_{k}} \right)
\end{equation}
which describes the evolution of the perturbed region.

During the radiation-dominated epoch the Universe can be regarded as a diluted gas with its Equation of State (EoS) written as $p=w\rho$ \citep[e.g.][]{2003LNP...631..301C,Ryden2003} where $p$ is pressure, $\rho$ is cosmological density, and the dimensionless quantity $w$ (the \emph{EoS parameter}) is equal to $\frac{1}{3}$ \citep[e.g.][]{2003LNP...631..301C}, since the sound speed is $c_{s}^{2}=\left(\frac{\partial p}{\partial \rho}\right)_{S}=w=\frac{1}{3}$ \citep[e.g.][]{1999PhRvD..59d3517S}. If during the QCD phase transition the Universe becomes matter-dominated (pressureless gas), then we get $w=0$ and $c_{s}^{2}=0$. Considering that $\rho(t)\propto a(t)^{-3(1+w)}$ \citep[e.g.][]{2012PhRvD..86a0001B} we write $\rho(\tau)=K_{s}s(\tau)^{-3(1+w)}$ where $K_{s}$ is a constant and $s(\tau)$ represents the evolution of the scale factor for the perturbed region. Similarly, $\rho_{k}=K_{k}a_{k}^{-3(1+w_{k})}$ with $w_{k}$ the EoS parameter when the fluctuation crosses the horizon. Hence, equation (\ref{CardallFuller_eq6a}) becomes
\begin{equation}
\label{CardallFuller_eq6a_sobrinho2008}
\left(\frac{ds}{d\tau}\right)^{2}=\frac{8\pi G}{3}\frac{K_{s}}{1+\delta_{k}} \left(
\frac{1+\delta_{k}}{s(\tau)^{1+3w}}-\frac{K_{k}}{K_{s}}\frac{\delta_{k}}{a_{k}^{1+3w_{k}}}
\right).
\end{equation}
The turnaround point ($t_c$) is reached when the perturbed region stops expanding, i.e., when $ds/d\tau=0$. Thus, evaluating equation (\ref{CardallFuller_eq6a_sobrinho2008}) at $t_c$ we get
\begin{equation}
\label{1998astro.ph..1103C_eq8_sobrinho2008}
s_{c}^{1+3w_{c}}=\frac{K_{s}}{K_{k}}a_{k}^{1+3w_{k}}\left(\frac{1+\delta_{k}}{\delta_{k}}\right)
\end{equation}
which relates the size of the perturbed region at the horizon crossing time with the respective size at the turnaround point. The calculation of the relation $K_{s}/K_{k}$ is detailed in Appendix \ref{appendixB}.

Following \citet[][]{1999PhRvD..59l4014J}, the value of $\delta_c$ during a first order phase transition is only a function of the horizon crossing time $t_k$ and the strength of the transition. We here interpret the latter as the time interval that the fluctuation spends on the mixed phase until the turnaround point ($t_c$) is reached. We are thus assuming, for our convenience, that when the turnaround point is reached a PBH will form, even if not immediately after the instant $t_c$ (as long as $\delta_k\geq \delta_{c}$). Hence, we are not taking into account the dynamics between the turnaround point and the instant when the PHB actually arises with the formation of an event horizon, which would require to numerically solve the Hernandez-Misner equations for a variable sound speed.

\subsection{The QCD phase transition according to three concurrent models}

A first order QCD phase transition is characterized by the formation of hadronic bubbles during a short period of supercooling when they grow slowly and the released latent heat maintains the temperature constant until the transition is complete \citep[e.g.][]{1999PhRvD..59d3517S}. We consider three concurrent approaches (models) to describe the QCD phase transition which we now describe in detail.

\subsubsection{The Bag Model (BM)}

The Bag Model (BM)  provides a first semiphenomenological description of an EoS that features a quark-hadron first-order transition \citep[e.g.][]{2006ARNPS..56..441B}. The simplest version of the model considers the thermodynamics in two different regions: a high temperature region ($T > T_{c}$, $t<t_{-}$) where we have a QGP and a low temperature region ($T < T_{c}$, $t>t_{+}$) where we have an HG. In both situations the sound speed is $c_{s}^{2}=1/3$. At the critical temperature $T = T_{c}$, i.e. during the time interval $[t_{-},t_{+}]$, quarks, gluons, and hadrons coexist in equilibrium at constant pressure and temperature and the sound speed vanishes \citep[e.g.][]{2006ARNPS..56..441B}.

\subsubsection{The Lattice Fit Model (LFM)}

In the limit, it is expected that simulations on larger and larger
lattices, while making the lattice spacing smaller and smaller, 
closely match the continuum theory. Here we consider a Lattice Fit Model (LFM) describing a first-order phase transition which was obtained for quenched lattice QCD (gluons only, no quarks), giving similar results to the case of two-flavour QCD  \citep[][]{1999PhRvD..59d3517S}.
However, now the high-temperature behaviour ($T\sim 4T_{c}\sim 700 \mathrm{~Mev}$) is not well described by the Stephan-Boltzmann law, which suggests that even at these high temperatures the plasma is not described by free quarks and gluons \citep[e.g.][]{2006ARNPS..56..441B}. Within the LFM the sound speed is given by \citep[][]{1999PhRvD..59d3517S}
\begin{eqnarray}
\label{Schmid1998_eq2.24}
c_{s}^{2}(T)=\frac{1}{3} \left(1-\frac{T_{c}}{T}\right)^{1-\gamma},& T\geq  T_{c}
\end{eqnarray}
with a good fit obtained for $0.3<\gamma<0.4$ (in this paper, we thus consider $\gamma=1/3$). Right after the end of the transition ($T<T_{c}$) an HG appears (similarly to the BM case)  and the sound speed is, once again, $c_{s}^{2}=1/3$ \citep[][]{1999PhRvD..59d3517S}.

\subsubsection{The scale factor and the average energy density of the Universe under the BM and the LFM}
\label{sec:The scale factor BM and LFM}

The evolution of the scale factor during the QGP and HG mixed phase in a first-order QCD transition, i.e., when $c_{s}^{2}=0$, is determined by the entropy conservation \citep[e.g.][]{2003AnP...515..220S, 1997PhRvL..78..791S}
\begin{equation}
\label{Delta_QCD_scale_factor}
f_{R}=\frac{a_{rl}(t_{+})}{a_{QCD}(t_{-})}=\left(1+R_{l}\left(\frac{g_{QGP}}{g_{HG}}-1\right)\right)^{1/3}
\end{equation}
with $R_{l}=1$ (BM) or $R_{l}=0.2$ (LFM). 

The evolution of the average energy density $\overline{\rho}$ as a function of time during a first-order QCD transition is given by \citep[][]{1997PhRvD..55.5871J}
\begin{eqnarray}
\label{Jedamzik1997_eq10}
\overline{\rho}(t)=\left(\frac{a_{QCD}(t_{-})}{a(t)}\right)^{3}\left(\rho_{QGP}(T_{c})\right.
\nonumber\\\left.+\frac{1}{3}\rho_{HG}(T_{c})\right)-\frac{1}{3}\rho_{HG}(T_{c}).
\end{eqnarray}
Taking into account that $\rho_{HG}(T)=\frac{\pi^{2}}{30}g_{HG}T^{4}$ and $\rho_{QGP}(T)= \frac{\pi^{2}}{30}g_{QGP}T^{4}+B$, with $B=\frac{\pi^{2}}{90}\left(g_{QGP}-g_{HG}\right)T_{c}^{4}$ \citep[cf.][]{1999PhRvD..59d3517S}  equation (\ref{Jedamzik1997_eq10}) becomes
\begin{equation}
\label{Jedamzik1997_eq10_sobrinho2007}
\overline{\rho}(t)=\frac{1}{3}\frac{\pi^{2}}{30}T_{c}^{4}\left[4g_{QGP}\left(\frac{a_{QCD}(t_{-})}{a(t)}\right)^{3}-g_{HG}\right].
\end{equation}
Let $\rho_{1}$ and $\rho_{2}$ ($\rho_{2}=\rho(t_{+})<\rho_{1}=\rho(t_{-})$) be the  energy densities at the start and at the end of the QCD phase transition, respectively. For simplicity, we assume for $\rho>\rho_{1}$ a pure QGP ($w=1/3$), for $\rho<\rho_{2}$ a pure HG ($w=1/3$) and for $\rho_{2}<\rho<\rho_{1}$ a mixed phase, that can be treated as dust ($w=0$). In order to characterize the horizon crossing time, $t_{k}$, and the turnaround point, $t_{c}$, in terms of energy density we introduce the quantities $x$ and $y$ defined as $x=\overline{\rho}_{k}/\rho_{1}$ ($\overline{\rho}_{k}$ represents the average cosmological density when $t=t_{k}$) and $y=\rho_{1}/\rho_{2}$. Taking into acount that $\rho_{1}=\rho(t_{-})=\rho_{QGP}(T_{c})$ we have, from equation (\ref{Jedamzik1997_eq10_sobrinho2007})
\begin{equation}
\label{xt_sobrinho2007}
x(t)=\frac{\overline{\rho}(t_{k})}{\rho_{1}}=\frac{4g_{QGP}\left(\frac{a_{QCD}(t_{-})}{a(t)}\right)^{3}-g_{HG}}{4g_{QGP}-g_{HG}}
\end{equation}
where $a(t)$ is given by:
(i) equation (\ref{scale factor radiation late}) if $x\leq y^{-1}$;
(ii) equation (\ref{scale factor dust QCD}) if $y^{-1}<x<1$;
(iii)  equation (\ref{scale factor radiation medium}) if $x\geq 1$.

\subsubsection{The Crossover Model (CM)}

An alternative to a first-order phase transition is a simple crossover in thermodynamic behaviour without discontinuities or singularities in the free energy or any of its derivatives. If the crossover is smooth the
system will evolve in local thermodynamic equilibrium. 
This is the situation if during the QCD phase transition only light quarks are present. Now, the sound speed decreases but does not vanish completely \citep[e.g.][]{2006ARNPS..56..441B}.
The entropy density for the QCD Crossover Model (CM) can be written as \citep[e.g.][]{1998MPLA...13.2771S}
\begin{eqnarray}
\label{Schmid1998_eq2.9}
S(T)=\frac{2\pi^{2}}{45}g_{HG}T^{3}\left[1+\frac{1}{2}\left(\frac{g_{QGP}}{g_{HG}} - 1\right)\right.\nonumber
\\
\left.\times\left(1+\tanh\left(\frac{T-T_{c}}{\Delta T}\right)\right)\right]
\end{eqnarray}
where the value of $\Delta T$ must be chosen in order to fit the modeled results: the best range of values found for the QCD is $0\leq \Delta T<0.1T_{c}$ \citep[e.g.][]{1999PhRvD..59d3517S,1997NuPhA.625..473B}. When $\Delta T\longrightarrow 0$ we recover the BM \citep[][]{2006ARNPS..56..441B}. We thus chose for the CM the values $\Delta T = 0.1T_{c}$ and $T_{c}=170$~MeV.

\section{The value of $\delta_{c}$ during the QCD phase transition}
\label{sec:delta_c QCD}

The relation between $a(t)$ and $T$ (the temperature of the Universe evaluated at the same instant) can be written as $a(t)=T_{0}/T$; in particular, when $t=t_{+}$ we have $a_{rl}(t_{+})=T_{0}/T_{c}$ (Table \ref{tabela-inicial}) which, together with equation (\ref{scale factor dust QCD}), gives
\begin{eqnarray}
\label{qcd_t_mais}
t_{+}=\frac{t_{eq}}{a_{m}(t_{eq})^{2}}\left(\frac{T_{0}}{T_{c}}\right)^{2}.
\end{eqnarray}
From equations (\ref{Delta_QCD_scale_factor}) and (\ref{scale factor dust QCD}) we get the relation
\begin{equation}
\label{qcd_t_menos}
t_{-}=\frac{t_{+}}{\sqrt{{f_{R}}^{3}}}.
\end{equation}
Using $T_{c}=170\mathrm{~MeV}$ (Table \ref{tabela-inicial}) we obtain, from equation (\ref{qcd_t_mais}), the value $t_{+}\approx 1.2\times10^{-4}\mathrm{~s}$ which is valid for both the BM and the LFM. From equation (\ref{qcd_t_menos}) we get $t_{-}\approx 6.2\times10^{-5}\mathrm{~s}$ (BM) or $t_{-}\approx 9.5\times10^{-5}\mathrm{~s}$ (LFM). The value of $y^{-1}$, which defines the end of the transition, can now be determined evaluating $x(t_{+})$. It turns out that for the BM case, with $g_{QGP}=61.75$ and $g_{HG}=17.25$, we get $y^{-1}\approx 0.225$ and, for the LFM,  $y^{-1}\approx 0.632$.

Fluctuation dynamics in the presence of a phase transition are dependent on the strength of the transition, as well as on the exact time $t_{k}$ at which the fluctuation crosses the horizon: shortly before onset, during, or shortly after completion of the transition \citep[][]{1999PhRvD..59l4014J}. For each situation we must also consider the possible locations of the turnaround point $t_{c}$ when the kinetic energy of expansion is zero and the region begins to collapse. 
Similarly to what was done by \citet[][]{1998astro.ph..1103C} we consider, in the case of a QCD first-order phase transition, six different classes of density fluctuations (designated A, B, C, D, E, and F). In each case the value of $\delta_c$ (cf. equation \ref{Bringmann_et_al_2001_eq5}) should de replaced by $\delta_{c}(1-f)$ where the function $f$ denotes the fraction of the overdense region spent in the dust-like phase of the transition (Appendix~\ref{appendixA} and  Table \ref{qcd fluctuations table}).

\subsection{The Bag Model (BM)}
\label{sec:The Bag Model}

\begin{figure}
\centering
\includegraphics[width=80mm]{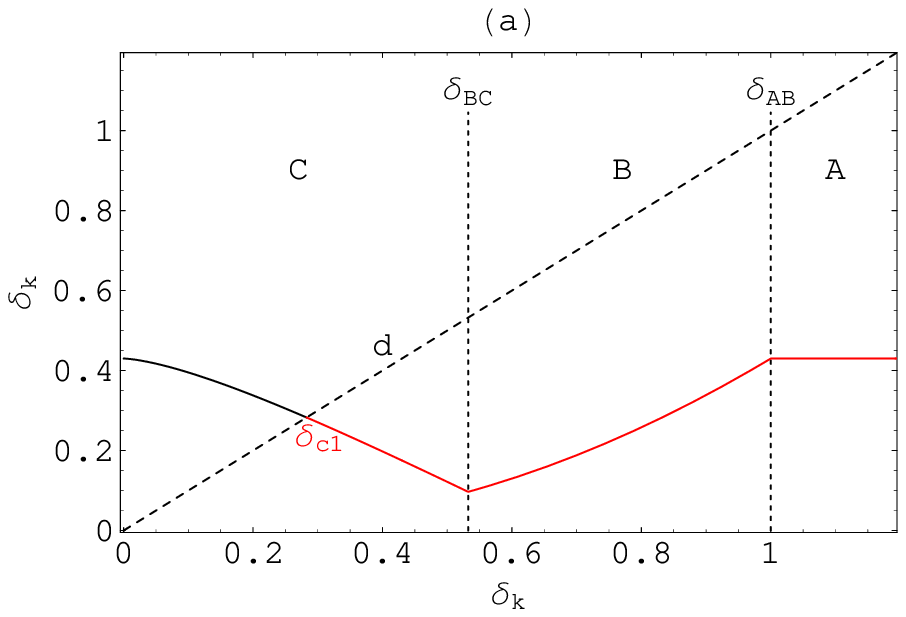} \\
\includegraphics[width=80mm]{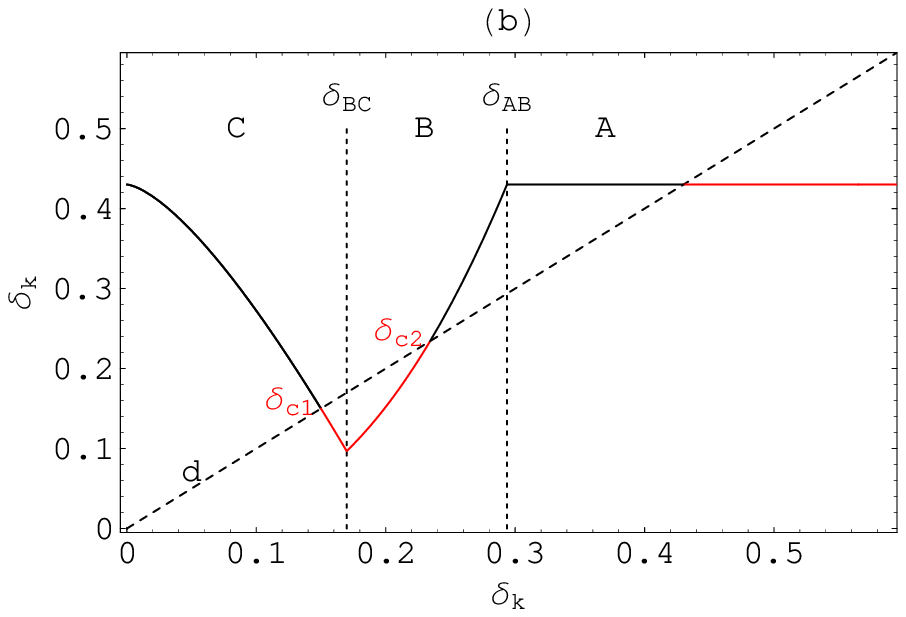}\\
\caption[]{The threshold for PBH formation during the QCD Bag Model (BM) when  $\delta_{c}=0.43$ and: {\bf (a)} $x=2$ ($t_{k}\approx 4.0\times10^{-5}\mathrm{~s}$); {\bf (b)} $x=15$ ($t_{k}\approx 1.1\times10^{-5}\mathrm{~s}$). The vertical dashed lines labeled ``$\delta_{AB}$" and ``$\delta_{BC}$" represent the frontiers between the different classes of fluctuations (A, B, and C), which cover the full range of $\delta_k$. The solid curve corresponds to the function $(1-f)\delta_{c}$ and the dashed curve, labeled ``$d$", to the identity $\delta_{k}$. A PBH will form if $(1-f)\delta_{c}<\delta_{k}$. In (a) this happens when $\delta_{k}>\delta_{c1}\approx 0.28$ (red curve). In (b) this happens when $\delta_{k}\in [\delta_{c1}\approx 0.15,\delta_{c2}\approx 0.23]$ and when  $\delta_{k}\geq 0.43$ (red curves). See text for more details.
\label{sobrinho2007_QCD_x2}}
\end{figure}

When $x> 1$ ($t_{k} < t_{-}$) we are dealing with fluctuations of classes A, B or C (Appendix~\ref{appendixA} and  Table \ref{qcd fluctuations table}). For a given $x$ we can determine the range of amplitudes which correspond to each class. For example, when $x=2$  ($t_{k}\approx 4.0\times10^{-5}\mathrm{~s}$) the overdensity will be of class C if $0<\delta_{k}<0.53$, of class B if $0.53<\delta_{k}<1$ and of class A if $\delta_{k}>1$. In order to identify the values of $\delta_{k}$ for which collapse to a PBH occurs, when $x=2$ and $\delta_{c}=0.43$, we  plot in Figure \ref{sobrinho2007_QCD_x2}a both $(1-f)\delta_{c}$ and $\delta_{k}$ itself as functions of $\delta_{k}$. It turns out that fluctuations of class C with $\delta_{k}< \delta_{c1}\approx 0.28$ dissipate before forming a PBH: $\delta_{c1}$ represents a lower threshold for PBH formation during the QCD phase transition.

As a second example we show in Figure \ref{sobrinho2007_QCD_x2}b the case $x=15$ ($t_{k}\approx 1.1\times10^{-5}\mathrm{~s}$). There are now two regions for which PBH formation is allowed:
(i) a region for $\delta_{k}\geq 0.43$, which corresponds to PBH formation from fluctuations of class A;
(ii) a region between $\delta_{c1}\approx 0.15$ and $\delta_{c2}\approx 0.23$  corresponding to the formation of PBHs from fluctuations of classes B and C. The gap between $\delta_{k}=0.23$ and $\delta_{k}=0.43$ corresponds to:
(i) fluctuations of class A which dissipate because they have $\delta_{k}<0.43$;
(ii) fluctuations of class B which dissipate because they do not spend enough time in the dust-like phase, allowing collapse to begin.

\begin{figure}
\centering
\includegraphics[width=84mm]{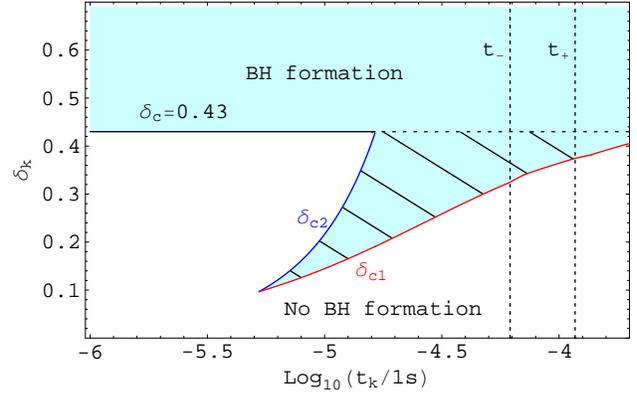}
\caption[]{The curve in the ($\log_{10}(t_{k}/\mathrm{1s}), \delta_{k}$) plane indicating which parameter values lead to collapse to a PBH under the QCD Bag Model (BM) when $\delta_{c}=0.43$. The vertical lines correspond to the beginning ($t_{k}=t_{-}$) and end ($t_{k}=t_{+}$) of the QCD phase transition. For a given horizon crossing time, $t_{k}$, the dashed region represents the newly found window for PBH formation 
(cf.~Figure~\ref{sobrinho2007_QCD_x2}). The lowest value attained by $\delta_{c}$ occurs at the intersection point $\delta_{c1}=\delta_{c2}=0.097$. It corresponds to a fluctuation crossing the horizon when $t_{k}=5.2\times10^{-6}\mathrm{~s}$.
\label{sobrinho2007_QCD_BagModel_deltac_t}}
\end{figure}

In Figure \ref{sobrinho2007_QCD_BagModel_deltac_t} we show the results obtained for the entire QCD Bag Model (BM) when $\delta_{c}=0.43$.
The intersection point $\delta_{c1}=\delta_{c2}=0.097$ represents the lowest value attained by $\delta_{c}$. It corresponds to a fluctuation that lies rigth on the boundary between fluctuations of classes B and C and crosses the horizon at the instant $t_{k}=5.2\times10^{-6}\mathrm{~s}$ (cf.~Figure~\ref{abcdef2015}). Depending on the instant of time that a particular fluctuation crosses the horizon we may have PBH forming from fluctuations of classes B, C, E and F, with the threshold $\delta_{c}<0.43$.

\subsection{The Crossover Model (CM)}
\label{sec:The Crossover}

We now assume that the minimum value of the sound speed  is reached at $t\approx t_{+}$ (corresponding to $T\approx T_{c}$) and that during the QCD the Universe continues to be radiation-dominated ($w=1/3$) with the scale factor given by equation (\ref{scale factor dust QCD}) with $n_{qcd}=1/2$. Considering that $a(T)=T_{0}/T$, we obtain 
\begin{eqnarray}
\label{temperature QCD crossover}
T(t)=T_{0}\left[a_{m}(t_{eq})\left(\frac{t}{t_{eq}}\right)^{1/2}\right]^{-1}.
\end{eqnarray}
Since the sound speed for a fluid with entropy density $S$ can be written as  \citep[e.g.][]{1999PhRvD..59d3517S}
\begin{equation}
\label{crossover-sound}
c_{s}^{2}=\left(\frac{d\ln S}{d \ln T}\right)^{-1}
\end{equation}
we get, from equation (\ref{Schmid1998_eq2.9})
\begin{eqnarray}
\label{sound_speed_crossover_qcd_sobrinho2007}
c_{s}^{2}(T)=\left[3+
\frac{T(g_{QGP}-g_{HG})}{\Delta T} \times\right.\nonumber\\\times
\left.
\frac{\mathrm{sech}\left(\frac{T-T_{c}}{\Delta T}\right)^{2}}{g_{HG}+g_{QGP}+ (g_{QGP}-g_{HG}) \mathrm{tanh}\left(\frac{T-T_{c}}{\Delta T}\right)}\right]^{-1}.
\end{eqnarray}

\begin{figure}
\centering
\includegraphics[width=84mm]{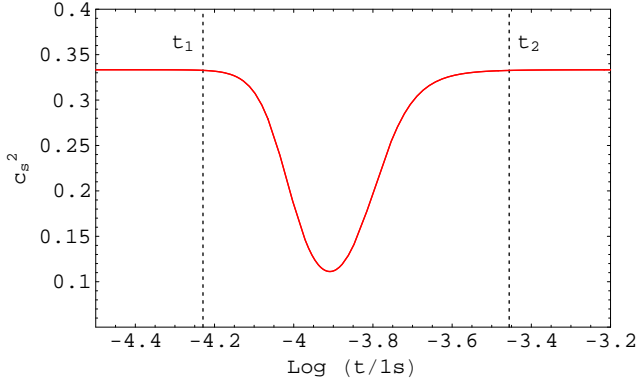}
\caption[]{The sound speed $c_{s}^{2}(t)$ for the QCD Crossover Model (CM) with a reference temperature $T_{c}=170\mathrm{~MeV}$ and $\Delta T=0.1 T_{c}$. Between the instants $t_{1}\approx 5.9\times10^{-5}\mathrm{~s}$ and $t_{2}\approx 3.5\times10^{-4}\mathrm{~s}$ the sound speed stays below its `background' value $c_{s,0}^{2}=1/3$, reaching a minimum of $c_{s,min}^{2}\approx 0.11$ at $t=t_{+}= 1.2\times10^{-4}\mathrm{~s}$.}
\label{sound_speed_crossover_time_2014}
\end{figure}

\begin{figure}
\centering
\includegraphics[width=84mm]{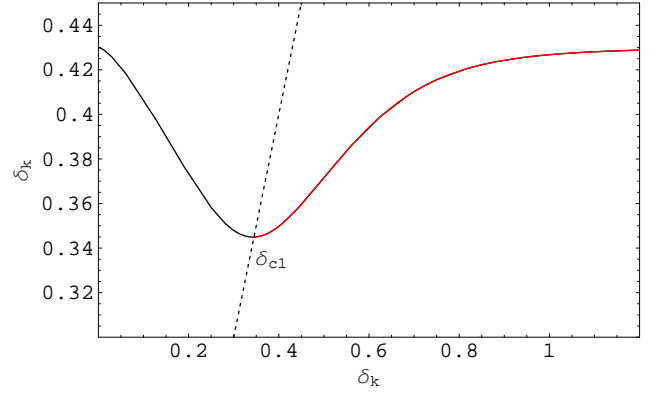}
\caption[]{PBH formation during the QCD transition according to the Crossover Model (CM) for the case $\delta_{c}=0.43$ and $t_{k}=4.2\times10^{-5}\mathrm{~s}$. The solid curve corresponds to the function $(1-f)\delta_{c}$ (cf. equation \ref{f_Crossover}) and the dashed curve corresponds to the identity $\delta_k$. The intersection between the two produces our newly found threshold for PBH formation ($\delta_{c1}\approx 0.345$). In red, we show where PBHs can form ($\delta_k>\delta_{c1}$). See the text for more details.}
\label{sobrinho2007_QCD_Crossover_antes++}
\end{figure}

\begin{figure}
\centering
\includegraphics[width=84mm]{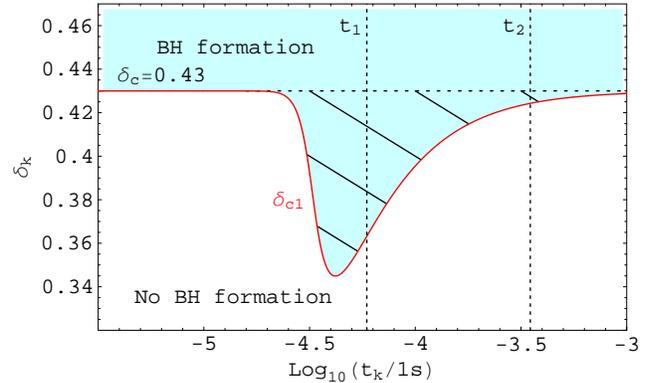}
\caption[]{The curve in the ($\log_{10}(t_{k}/\mathrm{1s}), \delta_{k}$) plane indicating which parameter values lead to collapse to a PBH for the QCD Crossover Model (CM) when  $\delta_{c}=0.43$. The vertical lines $t_{1}\approx 5.9\times10^{-5}\mathrm{~s}$ and $t_{2}\approx 3.5\times10^{-4}\mathrm{~s}$ define the transition epoch. For a given horizon crossing time, $t_{k}$, the dashed region represents our newly found window for PBH formation. 
\label{sobrinho2007_QCD_crossover_total}}
\end{figure}

We now define \textit{effective duration} as the time interval $[t_{1},t_{2}]$ for which the sound speed stays below the `background' value $c_{s,0}^{2}=1/3$. With the help of equations (\ref{temperature QCD crossover}) and (\ref{sound_speed_crossover_qcd_sobrinho2007}) and by taking $\Delta T=0.1T_{c}$ we get $t_{1}\approx 5.9\times10^{-5}\mathrm{~s}$ and $t_{2}\approx 3.5\times10^{-4}\mathrm{~s}$. In Figure \ref{sound_speed_crossover_time_2014}, we show the curve $c_{s}^{2}(t)$ with minimum $c_{s,min}^{2}\approx 0.11$. Now, taking $w_{k}=w_{c}=1/3$, equation (\ref{1998astro.ph..1103C_eq8_sobrinho2008}) can be written as
\begin{equation}
\label{Sc_Rh_Crossover}
s_{c}=a_{k}\left(\frac{1+\delta_{k}}{\delta_{k}}\right)^{1/2},
\end{equation}
which is valid for the entire transition. Expressions for $s_{c}$ and $a_{k}$ are obtained from the scale factor $a(t)$. Thus, we have
\begin{equation}
\label{QCD_crossover_Rh}
a_{k}=
a_{m}(t_{eq})\left(\frac{t_{k}}{t_{eq}}\right)^{1/2}
\end{equation}
and
\begin{equation}
\label{QCD_crossover_Sc}
s_{c}=
a_{m}(t_{eq})\left(\frac{t_{c}}{t_{eq}}\right)^{1/2}.
\end{equation}
Inserting equations (\ref{QCD_crossover_Rh}) and (\ref{QCD_crossover_Sc}) into equation (\ref{Sc_Rh_Crossover}) we obtain a relation between the  horizon crossing  time $t_{k}$ and the turnaround time $t_{c}$
\begin{equation}
\label{tk_tc_Crossover}
t_{c}=t_{k}\frac{1+\delta_{k}}{\delta_{k}}.
\end{equation}
We need to determine the analogous of function $f$  (Table \ref{qcd fluctuations table}) for the CM. This function should account for the fact that we now have a variable sound speed and that a smaller value of $c_{s}(t)$  contributes more significantly to the reduction of $\delta_{c}$ than a larger one. We then introduce the time function $\alpha_{sp}$ such that
\begin{equation}
\label{alpha_sound_sobrinho2007}
\alpha_{sp}(t)=1-\frac{c_{s}(t)}{c_{s0}}
\end{equation}
where $c_{s0}=1/\sqrt{3}$. In the case of the BM we have $\alpha_{sp}(t)=1$ during the mixed phase and $\alpha_{sp}(t)=0$ otherwise. Now, we consider for the function $f$ a more general equation
\begin{equation}
\label{f_general_sobrinho2007}
f=\frac{1}{s_{c}^{3}}\int_{s_{i}}^{s_{c}}\alpha_{sp}(t)ds^{3},
\end{equation}
where $s_{i}$ corresponds to the size of the region when the transition begins. Since $s=s(t)$ gives the evolution of the scale factor during the CM we get
\begin{equation}
\label{elementar_dS}
ds^{3}=\frac{3}{2}a_{m}(t_{eq})^{3}\frac{\sqrt{t}}{t_{eq}^{3/2}}dt.
\end{equation}
Combining equations (\ref{tk_tc_Crossover}), (\ref{alpha_sound_sobrinho2007}), (\ref{f_general_sobrinho2007}) and (\ref{elementar_dS}) we get
\begin{equation}
\label{f_Crossover}
f=\frac{3}{2}\left(t_{k}\frac{1+\delta_{k}}{\delta_{k}}\right)^{-3/2}\int_{t_{1}}^{t_{k}\frac{1+\delta_{k}}{\delta_{k}}}\left(1-\frac{c_{s}(t)}{c_{s0}}\right)\sqrt{t}dt.
\end{equation}
We can now study the changes in the value of the threshold $\delta_{c}$ during the CM following the same method that we used for the BM. In Figure \ref{sobrinho2007_QCD_Crossover_antes++} we show, as an example, the case $t_{k}=4.2\times 10^{-5}\mathrm{~s}$ giving $\delta_{c1}\approx 0.345$ (this is the case in which $\delta_{c1}$ reaches the smallest value).
Figure \ref{sobrinho2007_QCD_crossover_total} shows the region on the ($\log_{10}(t_{k}/\mathrm{1s}), \delta_{k}$) plane for which collapse to a PBH occurs in the case of the CM for $\delta_{c}=0.43$.

\subsection{The Lattice Fit Model (LFM)}
\label{sec:The Lattice Fit Model}

\begin{figure}
\centering
\includegraphics[width=84mm]{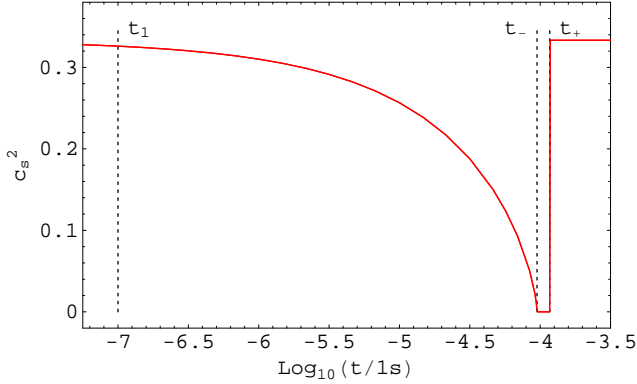}
\caption[]{The sound speed $c_{s}^{2}(t)$ for the QCD phase transition according to the Lattice Fit Model (LFM) with $T_{c}=170\mathrm{~MeV}$ (Table \ref{tabela-inicial}). During the mixed phase, which occurs between $t_{-}=9.5\times10^{-5}\mathrm{~s}$ and $t_{+}=1.2\times10^{-4}\mathrm{~s}$, the sound speed is zero. For $t_{1}\approx 10^{-7}\mathrm{~s}$ the sound speed equals 99\% of its `background' value of $1/\sqrt{3}$.}
\label{sound_speed_lattice_t_sobrinho2007}
\end{figure}

For the evolution of a fluctuation within the QCD Lattice Fit Model (LFM) we adopt a model similar to the one considered for the BM (cf. Appendix~\ref{appendixA}, Table~\ref{qcd fluctuations table}, Figure~\ref{abcdef2015}, Section~\ref{sec:The Bag Model}). One difference is that, in the case of the LFM, the mixed phase interval ($t_{-}<t<t_{+}$), during which the sound speed vanishes, is shorter. Another difference is that before the mixed phase (i.e. during the last instants of the QGP phase) there is a time interval $t_{1}\leq t\leq t_{-}$ during which the sound speed drops from $1/\sqrt{3}$ to zero. The sound speed as a function of time for $t\leq t_{-}$ can be written as (cf. equation \ref{Schmid1998_eq2.24})
\begin{equation}
\label{cs(t)_lattice}
c_{s}^{2}(t)=\frac{1}{3}\left(1-\frac{a_{rm}(t)}{a_{QCD}(t_{-})}\right)^{2/3}=\frac{1}{3}\left[1-\left(\frac{t}{t_{-}}\right)^{1/2}\right]^{2/3}
\end{equation}
We now define $t_{1}>t_{-}$ as the instant for which $c_{s}^{2}$ equals $99\%$ of its `background' value of $c_{s,0}^{2}=1/3$. From equation  (\ref{cs(t)_lattice}), with $t_{-}=~9.5\times~10^{-5}\mathrm{~s}$ we get $t_{1}\approx 10^{-7}\mathrm{~s}$ (see  Figure~\ref{sound_speed_lattice_t_sobrinho2007}).

We need to find a function $f$ suitable to the LFM. For the period $t_{1}\leq t\leq t_{-}$ we  proceed as in the CM (cf. Section \ref{sec:The Crossover}) while for the period $t_{-}\leq t\leq t_{+}$ we consider the BM results (cf. Section \ref{sec:The Bag Model}). Let us start with fluctuations of class A (see Table~\ref{qcd fluctuations table}). We  have $f_{A}=0$, as in the BM case, only if $t_{c}<t_{1}$. In general, for a fluctuation of class A we write (cf. equation~\ref{f_general_sobrinho2007})
\begin{equation}
\label{fA_lattice0}
f_{A_{LFM}}=\frac{1}{s_{c,A}^{3}}\int_{s_{k}}^{s_{c,A}}\alpha_{sp}(t)ds^{3}
\end{equation}
where $s_{c,A}$ is the size of the overdense region at turnaround (cf. Table \ref{qcd fluctuations table}) and $s_{k}= a_{rm}(t_{k})$ the size of the overdense region when the fluctuation crosses the horizon, given by equation (\ref{scale factor radiation medium}). Here $\alpha_{sp}(t)$ is given by equation (\ref{alpha_sound_sobrinho2007}), as in the CM, but now with the sound speed $c_{s}(t)$ given by equation (\ref{cs(t)_lattice}). Since the volume element $ds^{3}=3s^{2}ds=3s^{2}\frac{ds}{dt}dt$ must be evaluated in the radiation-dominated epoch ($t_{k}\leq t\leq t_{-}$), we have, from equation (\ref{scale factor radiation medium}), that
\begin{equation}
\label{elementar_dS2}
ds^{3}=\frac{3}{2}a_{QCD}(t_{-})^{3}\frac{\sqrt{t}}{(t_{-})^{3/2}}dt.
\end{equation}
Inserting equation (\ref{elementar_dS2}) into equation (\ref{fA_lattice0}) and using $a_{rm}(t_{k})$ from equation (\ref{scale factor radiation medium}) in $s_{c,A}$ (Table \ref {qcd fluctuations table}), we obtain
\begin{equation}
\label{fA_Lattice}
f_{A_{LFM}}=\frac{3}{2}\left(t_{k}\frac{1+\delta_{k}}{\delta_{k}}\right)^{-3/2}\int_{t_{k}}^{t_{k}\frac{1+\delta_{k}}{\delta_{k}}}\alpha_{sp}(t)\sqrt{t}dt.
\end{equation}
In the case of fluctuations of classes B and C (Table~\ref{qcd fluctuations table}) we write, respectively
\begin{eqnarray}
\label{fB_lattice_total}
f_{B}=f_{B_{LFM}}+\frac{s_{c,B}^{3}-s_{1}^{3}}{s_{c,B}^{3}}=\nonumber
\\=\frac{1}{s_{c,B}^{3}}\int_{s_{k}}^{s_{1}}\alpha_{sp}(t)ds^{3}+\frac{s_{c,B}^{3}-s_{1}^{3}}{s_{c,B}^{3}}
\end{eqnarray}
\begin{eqnarray}
\label{fC_lattice_total}
f_{C}=f_{C_{LFM}}+\frac{(s_{2})_{C}^{3}-s_{1}^{3}}{s_{c,C}^{3}}=\nonumber
\\=\frac{1}{s_{c,C}^{3}}\int_{s_{k}}^{s_{1}}\alpha_{sp}(t)ds^{3}+\frac{(s_{2})_{C}^{3}-s_{1}^{3}}{s_{c,C}^{3}}
\end{eqnarray}
where $s_{1}$ and $(s_{2})_{C}$ represent the size of the overdense region when, respectively, $t=t_{-}$ and $t=t_{+}$ (cf. equations \ref{S1} and \ref{S2C}) and $s_{c,B}$ and $s_{c,C}$ is the size of the overdense region at turnaround for, respectively, classes B and C (cf.~Table~\ref{qcd fluctuations table}). Inserting equation (\ref{elementar_dS2}) into equations (\ref{fB_lattice_total}) and (\ref{fC_lattice_total}), using $a_{rm}(t_{k})$ in $s_{c,B}$ and  $s_{c,C}$ and considering that $s_{1}= a_{QCD}(t_{-})$, we get
\begin{eqnarray}
\label{fB_Lattice}
f_{B_{LFM}}=\frac{3}{2}\left(t_{k}^{-1/2}x_{k}^{-1/4}\frac{(1+\delta_{k})^{3/4}}{\delta_{k}}\right)^{-3}
\nonumber\\
\times\int_{t_{k}}^{t_{-}}\alpha_{sp}(t)\sqrt{t}dt
\end{eqnarray}
\begin{equation}
\label{fC_Lattice}
f_{C_{LFM}}=\frac{3}{2}\left(t_{k}\frac{1+\delta_{k}}{\delta_{k}}\right)^{-3/2}y^{-1/2}\int_{t_{k}}^{t_{-}}\alpha_{sp}(t)\sqrt{t}dt.
\end{equation}

In the case of fluctuations of classes E and F ($t_{k}>t_{-}$) we continue to use, respectively for $f_{E}$ and $f_{F}$, the same expressions that we did in the case of the BM (Table \ref{qcd fluctuations table}).

\begin{figure}
\centering
\includegraphics[width=84mm]{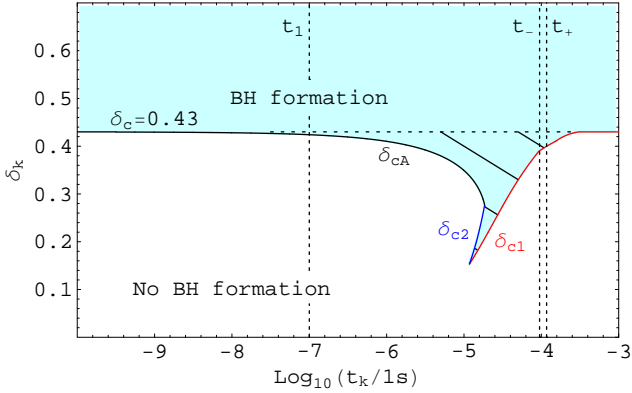}
\caption[]{The curve on the ($\log_{10}(t_{k}/1s)$, $\delta_{k}$) plane indicating which parameter values lead to collapse to a PBH in the case of the QCD Lattice Fit Model (LFM) when $\delta_{c}=0.43$. The vertical lines correspond to $t_{k}=t_{1}\approx 10^{-7}\mathrm{~s}$, $t_{k}=t_{-}=9.5\times10^{-5}\mathrm{~s}$, and $t_{k}=t_{+}=1.2\times10^{-4}\mathrm{~s}$ (cf. Figure \ref{sound_speed_lattice_t_sobrinho2007}). For a given horizon crossing time, $t_{k}$, the dashed region represents our newly discovered window for PBH formation. 
The lowest value attained by $\delta_{c}$ occurs at the intersection point $\delta_{c1}=\delta_{c2}=0.15$. It corresponds to a fluctuation crossing the horizon when 
$t_{k}=1.2\times10^{-5}\mathrm{~s}$.
\label{sobrinho2007_QCD_LatticeFit_deltac_t}}
\end{figure}

Following the same method that we used for the BM and the CM we have determined the threshold $\delta_{c}$ for the entire QCD phase transition according to the LFM. As a result, we obtain Figure \ref{sobrinho2007_QCD_LatticeFit_deltac_t}. 
Within the LFM our newly found window for PBH formation occurs between $\delta_{c1}$ and $\delta_{c}$,  and  $\delta_{c1}$ and $\delta_{c2}$; 
also, due to the decrease on the sound speed value between the instants $t_{1}$ and $t_{-}$ (cf. Figure \ref{sound_speed_lattice_t_sobrinho2007}) we have a new window for PBH formation from fluctuations of class A betwwen $\delta_{cA}$ and $\delta_{c}$. The intersection point $\delta_{c1}=\delta_{c2}=0.15$ represents the lowest value attained by $\delta_{c}$. It corresponds to a fluctuation that lies rigth on the boundary between fluctuations of classes B and C and crosses the horizon at the instant $t_{k}=1.2\times10^{-5}\mathrm{~s}$. Depending on the instant of time when a particular fluctuation crosses the horizon we might have PBH forming from fluctuations of classes A, B, C, E, and F, with the threshold $\delta_{c}<0.43$.

\subsection{Summary of results (the three models compared)}

We have determined the evolution ot the PBH formation threshold ($\delta_{c}$) during the QCD phase transition within three different models: Bag Model (BM), Crossover Model (CM), and Lattice Fit Model (LFM). 
Taking into account that the value of $\delta_{c}$ is related to the particular perturbation profile considered, we chose to work with $\delta_{c}=0.43$, corresponding to the Mexican-hat perturbation. This `background' value of $\delta_{c}$, which is valid for a radiation-dominated Universe, might decrease during the QCD phase transition, as a result of the reduction of the sound speed, improving the chances of PBH formation. 

In the case of the BM, we initially followed the literature (Section \ref{sec:The Bag Model}). Since the sound speed vanishes during the interval $[t_{-},t_{+}]\simeq[6.2\times10^{-5},1.2\times10^{-4}]$s, a new window for PBH formation appears (cf. Figure \ref{sobrinho2007_QCD_BagModel_deltac_t}) with the threshold for PBH formation reaching its absolute minimum value ($\delta_{c,min}$)  at the intersection point $\delta_{c1}=\delta_{c2}=0.097$.

For the CM (Section \ref{sec:The Crossover}) and the LFM (Section \ref{sec:The Lattice Fit Model}) we derived our own set of equations based on the behaviour of the sound speed during the QCD transition. Within the CM we found that the sound speed decreases smoothly to a minimum value of $c_{s}^{2}=0.11$, without vanishing, and rises again to its background value of $c_{s,0}^{2}=1/3$ (Figure \ref{sound_speed_crossover_time_2014}). Hence, a new window for PBH formation, between $\delta_{c1}$ and $\delta_{c}$, was discovered (cf. Figure \ref{sobrinho2007_QCD_crossover_total}) with the threshold for PBH formation reaching the minimum value $\delta_{c,min}=0.345$.

As regards the LFM model, the sound speed also vanishes during the interval $[t_{-},t_{+}]\simeq[9.5\times10^{-5},1.2\times10^{-4}]$s (as in the BM) while it is gradually reduced during the time interval $[t_{1},t_{-}]\simeq[10^{-7},9.5\times10^{-5}]$s  (cf. Figure \ref{sound_speed_lattice_t_sobrinho2007}). Hence, we treated the  interval $[t_{1},t_{-}]$ as in the CM and, the interval $[t_{-},t_{+}]$ as in the BM (cf. Section \ref{sec:The Lattice Fit Model}). As a result, we discovered, besides the new windows for PBH formation $[\delta_{c1},\delta_{c2}]$ and $[\delta_{c1},\delta_{c}]$ similar to the ones found on the BM, a hitherto unknown LFM window  for PBH formation [$\delta_{cA}$,$\delta_{c}$] (Figure~\ref{sobrinho2007_QCD_LatticeFit_deltac_t}), reflecting the contribution from the period [$t_{1}$,$t_{-}$] (cf.~Figure~\ref{sound_speed_lattice_t_sobrinho2007}). Within the LFM we found $\delta_{c,min}=0.15$.

In Table \ref{delta-c-min} we summarize the results obtained for $\delta_{c,min}$ during the QCD phase transition according to the BM (Figure \ref{sobrinho2007_QCD_BagModel_deltac_t}), CM (Figure \ref{sobrinho2007_QCD_crossover_total}), and LFM (Figure \ref{sobrinho2007_QCD_LatticeFit_deltac_t}). 
In the case of the BM we obtained a reduction of $\Delta\delta_{c}/\delta_{c}\approx 77\%$, while for the CM this is about 20\% and in the case of the LFM~65\%.

\begin{table}
\caption[]{The minimum value reached by the PBH formation threshold ($\delta_{c,min}$) within the Bag, Crossover, and Lattice Fit  Models (Figures \ref{sobrinho2007_QCD_BagModel_deltac_t}, \ref{sobrinho2007_QCD_crossover_total}, and~\ref{sobrinho2007_QCD_LatticeFit_deltac_t}, respectively) as well as the corresponding percentual decrease ($\Delta \delta_{c}/\delta_{c}$) with respect to the background value $\delta_{c}=0.43$. Also shown, in each case, is the corresponding instant of the horizon crossing time ($t_{k}$).\label{delta-c-min}}
\center
\begin{tabular}{cccc}
\hline
Model & $\delta_{c,min}$ & $\Delta \delta_{c}/\delta_{c}$(\%) & $t_{k}$~(s)\\
\hline
Bag (BM)& 0.097 & 77\% & $5.2\times10^{-6}$\\
Crossover (CM) & 0.345 & 20\% & $4.2\times10^{-5}$\\
Lattice Fit (LFM) & 0.15 & 65\% & $1.2\times10^{-5}$ \\
\hline
\hline
\end{tabular}
\end{table}

\section{Discussion}
\label{sec:discussion}

In order for the collapse of an overdense region in the early Universe forming a PBH, we must have $\delta_k > \delta_c$ where $\delta_k$ is the amplitude of the density fluctuation perturbation and $\delta_c$ a critical value which is related to the particular perturbation profile shape. We chose to work with $\delta_{c}=0.43$ since it corresponds to the representative Mexican-hat perturbation profile.
Although the value of $\delta_{c}$ remains constant during the radiation-dominated epoch, it can experience important reductions during cosmological phase transitions, which lead to a higher  probability $\beta(t_k)$ of forming PBHs. 

In particular, at the QCD phase transition $\sim 1$~M$_{\odot}$ PBHs might be formed \citep[][]{2010PhRvD..81j4019C}, even if only constituting a small fraction of the dark halo objets \citep[e.g.][]{2007A&A...469..387T}, they are not completely ruled out. 
A population of even smaller PBHs might also have formed, according to the scaling law given by equation (\ref{scaling-law}).

In this paper we explored the behaviour of $\delta_{c}$ during the QCD phase transition under three diferent models: Bag Model (BM), Crossover Model (CM), and Lattice Fit Model (LFM) and obtained reductions of  $20\%$ (CM),  $65\%$ (LFM), and $77\%$ (BM) -- cf.~Table~\ref{delta-c-min}.

Assuming, for example, $\sigma^{2}(t_{k})=2.3 \times 10^{-4}$ \citep[cf.][]{sobrinho2011} we get, for a radiation-dominated Universe 
($\delta_{c}=0.43$), $\beta(t_{k})\sim 10^{-177}$ (equation \ref{Bringmann_et_al_2001_eq5}), a negligible value. Our new results, however, show that $\delta_c$ can get as low as 0.097 (BM; Table~\ref{delta-c-min}), which implies the much higher probability of $\beta(t_k)\sim 10^{-10}$, which lies very close to the observational constraints at the QCD epoch \citep[cf.][]{2010PhRvD..81j4019C}. Although the BM is the one that offers the best prospects as regards PBH formation,  even in the case of the CM (a much smoother event: compare Figures \ref{sobrinho2007_QCD_BagModel_deltac_t} and \ref{sobrinho2007_QCD_crossover_total}) an important contribution to PBH formation is still expectable. Assuming, for example, $\sigma^{2}(t_{k})=2.0 \times 10^{-3}$ we get $\beta(t_k)\sim 10^{-14}$. 

Our next step will be to find an appropriate expression for the mass variance at horizon crossing, so that we might estimate the fraction of the Universe going into PBHs during the QCD epoch and, consequently, their cosmological density. This might be very relevant towards understanding the dark matter halo build-up in the Galaxy and in other galaxies.

\section*{Acknowledgements}

The authors are very grateful to Anne Marie Green for useful comments and observations made on this work. 
They also acknowledge helpful discussions with the referee  whose valuable comments improved this paper substantially.
All the Figures within this paper were created with the help of Wolfram Research, Inc., Mathematica, Version 5.1, Champaign, IL (2004).

\appendix

\section{Calculation of the relation $K_{s}/K_{k}$}
\label{appendixB}

Referring back to equation (\ref{1998astro.ph..1103C_eq8_sobrinho2008}) and to the constants $K_{s}$ and $K_{k}$, for classes A and F (see Appendix~\ref{appendixA} and Table~\ref{qcd fluctuations table}), which evolve completely during a radiation-dominated phase, and class D, which evolves completely during the mixed phase, we have $K_{s}/K_{k}=1$. In the case of class B, the change in the value of the EoS parameter (from $w_{k}=1/3$ to $w_{c}=0$) occurs when $t=t_{-}$, or equivalently, when $x=1$ ($x=\overline{\rho}_{k}/\rho_{1}$ with  $\overline{\rho}_{k}$ representing the average cosmological density when $t=t_{k}$ and $\rho_{1}$ the energy density at the start of the phase transition; cf.~Section~\ref{sec:The scale factor BM and LFM}). Considering that $\rho(\tau)\propto a(\tau)^{-3(1+w)}$  is a continuous function, we write, $K_{k}s_{1}^{-3(1+w_{k})}=K_{s}s_{1}^{-3(1+w_{c})}$, where $s_{1}$ represents the size of the overdense region at the beginning of the transition. This leads to
\begin{equation}
\label{ks/kk-classB}
\left( \frac{K_{s}}{K_{k}} \right)_B =\frac{s_{1}^{3w_{c}}}{s_{1}^{3w_{k}}}=\frac{1}{s_{1}}.
\end{equation}
In the case of class E the change of the EoS parameter (from $w_{k}=0$ to $w_{c}=1/3$) occurs when $t=t_{+}$ or, equivalently, when $x=y^{-1}$ ($y=\rho_{1}/\rho_{2}$ with $\rho_{2}$ representing the energy density at the end of the phase transition; cf. Section \ref{sec:The scale factor BM and LFM}). The continuity of the density $\rho$  now leads to
\begin{equation}
\label{ks/kk-classE}
\left( \frac{K_{s}}{K_{k}} \right)_E=\frac{s_{2}^{3w_{c}}}{s_{2}^{3w_{k}}}=(s_{2})_E
\end{equation}
where $s_{2}$ represents the size of the overdense region at the end of the transition.

In the case of fluctuations of class C there is an intermediate period when $w=w'=0$. Applying the continuity condition for $\rho$ successively at $t=t_{+}$ and $t=t_{-}$ we obtain
\begin{equation}
\label{ks/kk-classC}
\left( \frac{K_{s}}{K_{k}} \right)_C =\frac{s_{2}^{3w_{c}}}{s_{1}^{3w_{k}}}\frac{s_{1}^{3w'}}{s_{2}^{3w'}}=\frac{(s_{2})_C}{s_{1}}.
\end{equation}
The expression for $s_{1}$ (common to classes B and C) can be obtained considering that $\rho_{1}$ is reached from radiation domination (i.e. $\rho_{1}\propto  s_{1}^{-4}$ and  $\rho_{k}\propto  a_{k}^{-4}$). From energy conservation we have the condition $\rho_{1}s_{1}^{4}=\rho_{k}a_{k}^{4}$ which can be combined with equation (\ref{overdensity}) in order to obtain
\begin{equation}
\label{S1}
s_{1}=x^{1/4}(1+\delta_{k})^{1/4}a_{k}.
\end{equation}
An expression for $(s_{2})_E$ is obtained considering that $\rho_{2}$ is reached from the mixed phase (i.e. $\rho_{2}\propto  s_{2}^{-3}$ and  $\rho_{k}\propto  a_{k}^{-3}$). From energy conservation we have the condition $\rho_{2}s_{2}^{3}=\rho_{k}a_{k}^{3}$ which can be combined with equation (\ref{overdensity}) in order to obtain
\begin{equation}
\label{S2E}
(s_2)_E=(xy)^{1/3}(1+\delta_{k})^{1/3}a_{k}.
\end{equation}
An expression for $(s_{2})_C$ is obtained considering that a fluctuation of class C spans the entire mixed phase. From energy conservation we have the condition $\rho_{2}s_{1}^{3}=\rho_{k}s_{2}^{3}$ yielding
\begin{equation}
\label{S2C}
(s_2)_C=s_{1}y^{1/3}
\end{equation}
with $s_{1}$ given by equation (\ref{S1}). Finally, in the case of fluctuations of class F, which evolve completely during the radiation-domination phase, we have
\begin{equation}
\label{S2F}
(s_2)_F=(xy)^{1/4}(1+\delta_{k})^{1/4}a_{k}.
\end{equation}
The values here derived in equations (\ref{ks/kk-classB}-\ref{S2F}) are used to calculate the fraction of the overdense region spent in the dust-like phase of the QCD phase transition ($f$, in the last column of Table \ref{qcd fluctuations table}).

\section{Classes of fluctuations}
\label{appendixA}

Fluctuation dynamics in the presence of a phase transition are dependent on the strength of the transition, as well as on the horizon crossing time $t_{k}$ and on the turnaround point $t_{c}$ where the collapse begins \citep[][]{1999PhRvD..59l4014J}. In order to consider all possible solutions, we define six classes of density fluctuations labeled A, B, C, D, E, and F (Table~\ref{qcd fluctuations table}). In each case the value of the PBH formation threshold $\delta_c$ (cf. equation \ref{Bringmann_et_al_2001_eq5}) must de replaced by $\delta_{c}(1-f)$ where $f$ denotes the fraction of the overdense region spent in the dust-like phase of the transition.

\begin{table*}
\caption[]{Classification of overdense regions according to the state of matter at the horizon crossing time ($t_{k}$) and at the turnaround point ($t_{c}$) for the QCD phase transition. We are using a convention first proposed by \citet[][]{1998astro.ph..1103C}. {\bf (1)}: fluctuation class (class D fluctuations are shown only for the sake of completeness, since they are ruled out in our study --- we are looking for values of $\delta_{c}$ that are always less or equal to 0.43 and class D fluctuations do not exist when $\delta_{k}<1.55$ (cf. equation \ref{classCclassD_single_point})); {\bf (2)}: the state of matter at the horizon crossing time ($t_{k}$); HG --- Hadronic Gas; QGP --- Quark-Gluon Plasma; mixed --- mixture of HG and QGP; {\bf (3)}: EoS parameter at horizon crossing; $w=1/3$ --- radiation-dominated Universe; $w=0$ --- matter-dominated Universe; {\bf (4)}: the state of matter at the turnaround point ($t_{c}$); {\bf (5)}: EoS parameter at turnaround; {\bf (6)}: the scale factor of the overdense region at turnaround (cf. equation \ref{1998astro.ph..1103C_eq8_sobrinho2008}) --- $x=\overline{\rho}_{k}/\rho_{1}$ and $y=\rho_{1}/\rho_{2}$ ($\overline{\rho}_{k}$ represents the average cosmological density when $t=t_{k}$, $\rho_{1}$ and $\rho_{2}$ are the energy densities at the start and at the end of the phase transition, respectively), $a_{k}$ is the size of the overdense region when $t=t_{k}$; {\bf (7)}: the fraction of the overdense region spent in the dust-like phase of the transition -- $s_{1}$ and $s_{2}$ represent the size of the overdense region at the start and at the end of the phase transition, respectively (cf.~equations~\ref{S1}--\ref{S2F}). \label{qcd fluctuations table}}
\center
\begin{tabular}{ccccccc}
\hline
\hline
\\
{\bf (1)} & {\bf (2)} & {\bf (3)} & {\bf (4)} & {\bf (5)} & {\bf (6)} & {\bf (7)} \\
\\
Class  & $t_k$  & $w_k$ &  $t_c$ & $w_c$ & $s_{c}$ & $f$ \\
\\
\hline
\\
A & QGP & $1/3$ & QGP & $1/3$ & $a_{k}\left(\frac{1+\delta_{k}}{\delta_{k}}\right)^{1/2}$ & 0\\
\\
B & QGP & $1/3$ & mixed & 0 & $a_{k}\frac{x^{-1/4}(1+\delta_{k})^{3/4}}{\delta_{k}}$ & $\frac{s_{c,B}^{3}-s_{1}^{3}}{s_{c,B}^{3}}$\\
\\
C & QGP & $1/3$ & HG & $1/3$ & $a_{k}y^{1/6}\left(\frac{1+\delta_{k}}{\delta_{k}}\right)^{1/2}$ & $\frac{(s_{2})_{C}^{3}-s_{1}^{3}}{s_{c,C}^{3}}$\\
\\
{\it D} & {\it mixed} & {\it 0} & {\it mixed} & {\it 0} & $a_{k}\frac{1+\delta_{k}}{\delta_{k}}$ & $\frac{s_{c,D}^{3}-s_{1}^{3}}{s_{c,D}^{3}}$\\
\\
E & mixed & 0 & HG & $1/3$ & $a_{k}(xy)^{1/6}\frac{(1+\delta_{k})^{2/3}}{\delta_{k}^{1/2}}$ & $\frac{(s_{2})_{E}^{3}-s_{1}^{3}}{s_{c,E}^{3}}$ \\
\\
F &  HG & $1/3$ &  HG & $1/3$   & $a_{k}\left(\frac{1+\delta_{k}}{\delta_{k}}\right)^{1/2}$ & $\frac{(s_{2})_{F}^{3}-s_{1}^{3}}{s_{c,F}^{3}}$\\
\hline
\end{tabular}
\end{table*}

The boundary between two neighbouring classes X and Y can be determined taking into account that it occurs where $\delta_{c}(1-f_{X})=\delta_{c}(1-f_{Y})$. For example, for the separation between classes A and B we get the condition $x(t_{k})=(1+\delta_{k})/\delta_{k}^{2}$ \citep[see][for a detailed discussion]{sobrinho2011}.

The boundary between classes C and D turns out to be a single point
\begin{equation}
\label{classCclassD_single_point}
y=\left(\frac{1+\delta_{k}}{\delta_{k}}\right)^{3} \approx  \frac{1}{0.225} \Rightarrow \delta_k \approx  1.55
\end{equation}
meaning that class D fluctuations do not exist for $\delta_{k}<1.55$ and class C ones for $\delta_{k}>1.55$. Since, in our study, we are looking for values of $\delta_{c}$ that are always less or equal to 0.43, class D is thus ruled out.

\begin{figure}
\centering
\includegraphics[width=84mm]{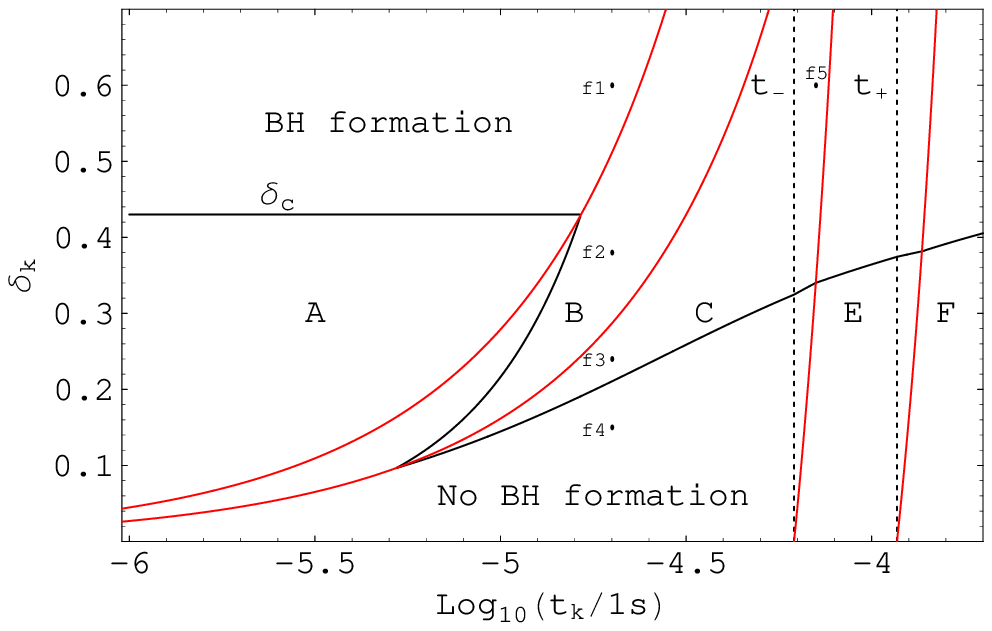}
\caption[]{ 
Regions in the plane ($\log_{10}(t_{k}/1\mathrm{s}),\delta_{k}$) corresponding to the classes of fluctuations listed in Table~\ref{qcd fluctuations table} for the BM case: the black line labeled ``$\delta_c$" represents the PBH formation threshold during the QCD phase transition according 
to this model(Section~\ref{sec:The Bag Model}, Figure~\ref{sobrinho2007_QCD_BagModel_deltac_t}). 
The red lines split the plane into slices, each one corresponding to a different class of fluctuations --- A, B, C, E, or F (D is not represented because this class only exists for $\delta_{k}>1.55$ --- cf. equation \ref{classCclassD_single_point}, Table~\ref{qcd fluctuations table}). The points f1, f2, f3, and f4 represent fluctuations with different amplitudes $\delta$ that cross the horizon at the same instant $t_{k}=2\times10^{-5}\mathrm{~s}$ during the QGP phase. The point $f5$ represents a fluctuation which crosses the horizon at the instant $t_{k}=7\times10^{-5}\mathrm{~s}$.
These five points are used as an explanatory example in the text.
\label{abcdef2015}}
\end{figure}

The use of six different classes of fluctuations works as an intermediate step towards determining the evolution of $\delta_c$ during the QCD phase transition. In Figure~\ref{abcdef2015} we reproduce Figure~\ref{sobrinho2007_QCD_BagModel_deltac_t} but now including very detailed explanations on five fluctuation classes (D is not represented --- cf. equation~\ref{classCclassD_single_point}). In it, the points f1, f2, f3, and f4, for example, represent fluctuations with increasing amplitude $\delta$ that cross the horizon at the same instant $t_{k}=2\times10^{-5}\mathrm{~s}$ before the beginning of the QCD phase transition.

We now use these four example points by following the details in both Figure~\ref{abcdef2015} and Table~\ref{qcd fluctuations table}.
f1, the point with the largest amplitude ($\delta$) develops faster, reaching the turnaround point (not represented) before $t_{-}$. Therefore it is classified as a fluctuation of class A. f2 represents a fluctuation of class B: although it crosses the horizon before $t_{-}$, the turnaround point is reached only during the mixed phase because now we have a fluctuation with a smaller amplitude which develops slower. f3 represents a fluctuation of class C: it starts before $t_{-}$ and ends after $t_{+}$, crossing the all phase transition in between. In the case of f1, f2, and f3 we have PBH formation. On the other hand, f4 represents a fluctuation with amplitude below the threshold for PBH formation. It crosses the QCD phase transition and dissipates without reaching a turnaround point.

In the case of an unperturbed region the beginning of the QCD phase transition occurs at $t_{-}$ and the end at $t_{+}$. If the region is 
perturbed, however, the evolution is slower. 
For example, when $t_{k}=7\times10^{-5}\mathrm{~s}$, although the average cosmological background is already on the mixed phase ($t_{-}<t_{k}<t_{+}$), an overdense region with, for example, $\delta_{k}=0.6$ remains on the QGP phase --- e.g. point f5 in Figure~\ref{abcdef2015}: such regions do not expand with the rest of the background Universe (the state of matter on those regions evolves slower).

\end{document}